\documentclass[conference]{IEEEtran}
\IEEEoverridecommandlockouts
\usepackage[dvips]{psfrag}
\usepackage[dvips]{epsfig}
\usepackage{epic,color}
\usepackage[cmex10]{amsmath}
\usepackage{amsfonts}
\usepackage{amssymb}
\usepackage{amsmath}
\usepackage{yfonts}
\usepackage{array}
\usepackage{bigstrut}
\usepackage{multirow}
\usepackage{dsfont}
\usepackage{cite}
\usepackage{fixltx2e}
\usepackage{accents}
\usepackage{mathtools}
\usepackage[normalem]{ulem}
\usepackage{enumerate}
\usepackage{stfloats}
\usepackage[T1]{fontenc}
\usepackage{float}
\usepackage{epstopdf}
\usepackage{algorithm}
\usepackage{algorithmic}
\usepackage{caption}
\usepackage{acronym}
\usepackage[caption=false]{subfig}
\usepackage{breqn}
\usepackage{xparse}% http://ctan.org/pkg/xparse
\usepackage[T1]{fontenc}
\usepackage{calligra,amsmath,amssymb}

\NewDocumentCommand{\ceil}{s O{} m}{%
  \IfBooleanTF{#1} % starred
    {\left\lceil#3\right\rceil} % \ceil*[..]{..}
    {#2\lceil#3#2\rceil} % \ceil[..]{..}
}
\NewDocumentCommand{\floor}{s O{} m}{%
  \IfBooleanTF{#1} % starred
    {\left\lfloor#3\right\rfloor} % \floor*[..]{..}
    {#2\lfloor#3#2\rfloor} % \floor[..]{..}
}

\newtheorem{lemma}{Lemma}

\newcommand{\mbf}[1]{\mathbf{#1}}
\newcommand{\ub}[1]{\underbrace{#1}}

\usepackage[keeplastbox]{flushend}

\usepackage{color}

\acrodef{CRN}{cognitive radio network}
\acrodef{SU}{secondary user}
\acrodef{PU}{primary user}
\acrodef{ZF}{zero forcing}
\acrodef{FD}{full-duplex}
\acrodef{BS}{micro base station}
\acrodef{i.i.d.}{independent and identically distributed}
\acrodef{DL}{downlink}
\acrodef{UL}{uplink}
\acrodef{SINR}{signal-to-interference noise ratio}
\acrodef{SNR}{signal noise ratio}
\acrodef{AWGN}{additive white Gaussian noise}
\acrodef{MMSE}{minimum mean square error}
\acrodef{SIC}{successive interference cancellation}
\acrodef{SI}{self-interference}
\acrodef{CCI}{co-channel interference}
\acrodef{MUI}{multiuser interference}
\acrodef{NOMA}{non-orthogonal multiple access}
\acrodef{OMA}{orthogonal multiple access}
\acrodef{QoS}{quality of service}
\acrodef{SIC}{successive interference cancellation}
\acrodef{SVD}{singular value decomposition}
\acrodef{MIMO}{multiple-input and multiple-output}
\acrodef{SISO}{single-input and single-output}

\begin{document}

\title{A Fair Individual Rate Comparison between MIMO-NOMA and MIMO-OMA}

%\author{\IEEEauthorblockN{Ming Zeng, \emph{Student Member, IEEE},
%Animesh Yadav, \emph{Member, IEEE}, Octavia A. Dobre, \emph{Senior Member}, \emph{IEEE}
%}
\author{\IEEEauthorblockN{Ming Zeng,
Animesh Yadav, Octavia A. Dobre and H. Vincent Poor\IEEEauthorrefmark{1} }
\IEEEauthorblockA{Faculty of Engineering and Applied Science, Memorial University, St. John, Canada \\
\IEEEauthorrefmark{1}Department of Electrical Engineering, Princeton University, Princeton, NJ, USA}
%\IEEEauthorblockA{\IEEEauthorrefmark{2}School of Electrical and Computer Engineering, National Technical University of Athens, Greece}
%\IEEEauthorblockA{\IEEEauthorrefmark{3}Faculty of Engineering and Applied Science, Memorial University, St. John, Canada}

\IEEEauthorblockA {Email: \{mzeng, animeshy, odobre\}@mun.ca, poor@princeton.edu\IEEEauthorrefmark{1} }}
%, , \\Georgios I. Tsiropoulos, \emph{Member, IEEE}, and H. Vincent Poor, \emph{Fellow, IEEE}

%\author{\noindent \begingroup\centering{Ming Zeng,~\IEEEmembership{Student Member,~IEEE}, and Others, \IEEEmembership{Senior Member, IEEE}}\endgroup%
%\thanks{}}
\maketitle
\begin{abstract}
In this paper, we compare the individual rate of MIMO-NOMA and MIMO-OMA when users are paired into clusters. A power allocation (PA) strategy is proposed, which ensures that MIMO-NOMA achieves a higher individual rate for each user than MIMO-OMA with arbitrary PA and optimal degrees of freedom split. In addition, a special case with equal degrees of freedom and arbitrary PA for OMA is considered, for which the individual rate superiority of NOMA still holds. Moreover, it is shown that NOMA can attain better fairness through appropriate PA. Finally, simulations are carried out, which validate the developed analytical results.

%Compared with existing individual rate comparison schemes, which assign equal power and degrees of freedom to users in the same cluster for OMA, the considered scheme is general and fair, since the power allocation in each cluster for OMA can be arbitrary, and the degrees of freedom is split such that OMA maximizes its sum rate. In addition, a special case with equal degrees of freedom and arbitrary power allocation for OMA is also considered, and the conclusion is shown to still hold. Finally, simulations are carried out, which validate the developed analytical results.

%and conduct comparison between NOMA and OMA in two scenarios. The first scenario consists of two user with equal DoF, in which the weak user has a minimum rate requirement. We prove that NOMA always achieves a higher sum rate than OMA. The second scenario considers the more general case, in which the DoF is an optimization variable. Numerical study shows that NOMA achieves a higher sum rate than OMA. 

%first provide the condition of obtaining optimal capacity for OMA. Based on this, we show that NOMA still outperforms OMA for two MIMO scenarios: a) two user MIMO; and b) amulti-user MIMO with two users paired into a cluster. Moreover, we prove that each user in NOMA can obtain a larger data rate than its counterpart through appropriate power allocation. 
\end{abstract}

%H. V. Poor is with Princeton University, Princeton, NJ 08544 USA (e-mail: poor@princeton.edu).}  }

%\IEEEpeerreviewmaketitle{\noindent }
%\begin{IEEEkeywords}
%\noindent
%\end{IEEEkeywords}

\section{Introduction}
The non-orthogonal multiple access (NOMA) has
drawn great attention as a promising
technology for improving the spectral efficiency for the next generation mobile communication networks \cite{22, 3, 2, 16, 12, 24}. There exist two main NOMA schemes, i.e., power-domain and code-domain NOMA. In this paper, we focus on the former, in which users are multiplexed on power domain. For notational simplicity, we refer to power-domain NOMA as NOMA.

A few works have verified via simulation the superiority of NOMA over OMA for multi-user scenarios in term of achievable sum rate \cite{1, 9, 5, 8}. For single-input single-output (SISO) systems, \cite{1} shows that NOMA can achieve a larger sum rate, while \cite{9} illustrates that a larger ergodic sum rate is obtained by NOMA for a cellular downlink system with randomly deployed users. As for multiple-input multiple-output (MIMO) systems, \cite{5,8} provide some insight: \cite{5} verifies that a larger ergodic sum rate for two users can be obtained by NOMA, whereas \cite{8} shows that NOMA can achieve a larger sum rate for a multi-user scenario, with two users paired into a cluster, and sharing a common transmit beamforming vector.

Some recent works aim to analytically prove that NOMA achieves higher sum rate over OMA. For SISO systems, power allocation (PA) in \cite{20} is conducted to guarantee that NOMA achieves a larger sum rate than OMA with equal power coefficients and degrees of freedom (DoF). For MIMO systems, \cite{6} derives the sum rate gain of NOMA over OMA under two extreme cases of user pairing: 1)  the best user with the worst user; 2) the best user with the second best user. Moreover, a cognitive radio inspired PA is proposed, which ensures that the data rate of the weak user is larger than that in OMA. However, the sum rate for OMA is not optimized in the above works as equal power and DoF are allocated to users. In \cite{18,21}, the authors overcome this issue, and demonstrate that NOMA achieves a larger sum rate than OMA for scenarios with two users and multiple users per cluster, respectively.

The major drawback of the sum rate comparison is that it neglects fairness. To the best of our knowledge, none of the previous works considers fairness during sum rate comparison. Note that although simulation results in \cite{21} show that NOMA achieves higher fairness, no theoretical analysis is provided. Hence, in order to account for fairness, we need to compare the individual rates of the users. In particular, the individual rate for any user in NOMA should be higher or equal than its counterpart in OMA. In \cite{17}, the PA scheme for a SISO system is designed such that the individual rate of each user in NOMA is guaranteed to be larger than its counterpart in OMA. However, \cite{17} still adopts equal PA and DoF for OMA, which is suboptimal. By filling in this gap, the main contribution of this paper lies in:
\begin{itemize}
\item A general and fair individual rate comparison is considered, in which the PA for OMA is arbitrary and the DoF is split such that the {\color{black}maximum sum rate} in OMA is achieved. On this basis, a PA strategy is proposed, which ensures that NOMA achieves higher individual rates than OMA. 

\item For the particular case with equal DoF and arbitrary PA, analytical results are provided to demonstrate the superiority of NOMA over OMA in terms of individual rates. 

\item In addition to the individual rate superiority, it is also shown that better fairness is achieved by NOMA through appropriate PA. 
\end{itemize}
 
%To address this, we consider a general and fair individual rate comparison scheme in this paper, which allows the PA in OMA to be arbitrary, and the DoF is split such that the {\color{black}capacity} in OMA is achieved. On this basis, we propose a PA strategy, which ensures that NOMA achieves higher individual rate than OMA. In addition, for a special case with equal DoF, analytical results are also developed to demonstrate the superiority of NOMA over OMA in terms of individual rate. 

The rest of the paper is organized as follows. The system model is introduced in Section II. The individual rate comparison between MIMO-NOMA and MIMO-OMA is conducted in Section III, where a PA strategy is additionally proposed. The particular case of equal DoF is also discussed in Section III, while simulation results are shown in \text{Section IV}. Conclusions are finally drawn in Section V.

\section{System Model and Problem Formulation}
\subsection{System Model}
A multi-user MIMO-NOMA downlink transmission scenario is investigated, in which a \ac{BS} deployed with $M$ antennas sends information to $2M$ users, each with $N$ antennas. Two users are paired into a cluster for complexity reduction \cite{6}, and NOMA is only applied between them. Accordingly, there are $M$ clusters in the system. We adopt the block fading channel model, where both path loss component and small scale fading are considered, e.g., the channel matrix from the \ac{BS} to user $k, k\in\{1,2\}$ in cluster $m, m\in\{1,\ldots, M\}$, is $\mbf{H}_{m,k}=\mbf{G}_{m,k}/{L_{m,k}}$, with $\mbf{G}_{m,k}\in \mathbb{C}^{N\times M}$ denoting the Rayleigh fading channel matrix and $L_{m,k}$ representing the path loss component. The transmit and receive beamforming vectors fulfill the following conditions \cite{8}: 1) zero-forcing (ZF) precoding is conducted at the \ac{BS} to remove the inter-cluster interference; 2) signal alignment is conducted at the receiver between users in the same cluster, i.e., ${\mbf{v}_{m,2}^H}\mbf{G}_{m,2} = {\mbf{v}_{m,1}^H}\mbf{G}_{m,1}$, where $\mbf{v}_{m,k}^H$ denotes the receive beamforming vector.

%The \ac{BS} multiplexes users in the power domain and multiplies the transmit signal by a precoding matrix $\mbf{P}\in \mathbb{C}^{M\times 2Q}$ before transmission. Thus, 

As users in the same cluster share a common transmit beamforming vector, the signal transmitted from the \ac{BS} can be expressed as
\begin{equation}
\mbf{x}=\mbf{P}{\mbf{s}},
\end{equation}
where $\mbf{P}=[\mbf{p}_{1} ~\cdots ~ \mbf{p}_{M}] \in \mathbb{C}^{M\times M}$, with $\mbf{p}_{m} \in \mathbb{C}^{M\times 1}$ representing the normalized transmit beamforming vector for cluster $m$. Additionally, the information bearing vector ${\mbf{s}}\in \mathbb{C}^{M\times 1}$ is given by
\begin{equation}
{\mbf{s}}=
\begin{bmatrix}
\alpha_{1,1}s_{1,1}+\alpha_{1,2}s_{1,2} \\
\vdots \\
\alpha_{M,1}s_{M,1}+ \alpha_{M,2}s_{M,2} 
\end{bmatrix},
\end{equation}
where $s_{m,k}$ and $\alpha_{m,k}$ represent the signal and corresponding PA coefficient for user $(m,k)$, respectively, satisfying $\alpha_{m,1}^2+\alpha_{m,2}^2=1, \forall m$. 

At the receiver of user $(m,k)$, the normalized receive beamforming vector $\mbf{v}_{m,k}$ is applied, and thus, the received signal $\mbf{y}_{m,k}$ is given by
\begin{IEEEeqnarray*}{lll}\label{eq:multi_cluster_Rx}
{\mbf{v}_{m,k}^H}\mbf{y}_{m,k}=&\alpha_{m,1}{\mbf{v}_{m,1}^H}\mbf{H}_{m,1}\mbf{p}_{m}s_{m,1}
+\alpha_{m,2}{\mbf{v}_{m,2}^H}\mbf{H}_{m,2}\mbf{p}_{m}s_{m,2} \nonumber \\
&+ \ub{ \displaystyle\sum^{M}_{i\neq m}\sum^2_{l=1}\alpha_{i,l}{\mbf{v}_{m,l}^H}\mbf{H}_{m,l}\mbf{p}_{i}s_{i,l}
}_{\text{interference from other clusters}}
+ {\mbf{v}_{m,k}^H}\mbf{n}_{m,k},\IEEEyesnumber
\end{IEEEeqnarray*}
where $(\cdot)^H$ represents the Hermitian transpose operation and $\mbf{n}_{m,k}\in \mathbb{C}^{N\times 1} \sim \mathcal{CN} (0,\sigma^2_{n}\mbf{I})$ is the \ac{AWGN} vector at user $(m,k)$.

As ZF precoding is adopted at the BS, inter-cluster interference can be eliminated, and thus, the cluster index $m$ can be dropped for notational simplicity. Consequently, the received signal can be rewritten as 
\begin{equation}
{\mbf{v}_{k}^H}\mbf{y}_{k}=\alpha_{1}{\mbf{v}_{1}^H}\mbf{H}_{1}\mbf{p}s_{1}
+\alpha_{2}{\mbf{v}_{2}^H}\mbf{H}_{2}\mbf{p}s_{2}
 + {\mbf{v}_{k}^H}\mbf{n}_{k}.
\end{equation}

Without loss of generality, the effective channel gains of the users are ordered as follows:
\begin{equation} \label{eq:order}
|{\mbf{v}_{1}^H}\mbf{H}_{1}\mbf{p}|^2 \geq  |{\mbf{v}_{2}^H}\mbf{H}_{2}\mbf{p}|^2.
\end{equation}

Accordingly, \ac{SIC} is conducted at user $1$ to remove the interference from user $2$, and because of this, the achieved data rate at user $1$ can be expressed as \cite{18}
\begin{equation} \label{eq:user1}
R_{1}^{\text{NOMA}}=\log_2
(1+  \rho  \alpha_{1}^2 |{\mbf{v}_{1}^H}\mbf{H}_{1}\mbf{p}|^2),
\end{equation}
where {\color{black} $\rho=\frac{1}{\mathbb{E} [|\mbf{v}_{k}^H \mbf{n}_{k}|^2  ] }$} is the same for the two users, as the receive beamforming vector is normalized and the noise variance remains unchanged after rotation. $\mathbb{E} [\cdot]$ denotes the expectation operator.

In contrast, user $2$ considers user 1's signal as interference, and thus, its achievable rate is given by
\begin{equation}
R_{2}^{\text{NOMA}}=\log_2 \label{eq:user2}
\bigg(1+\frac{\rho \alpha_{2}^2 |{\mbf{v}_{2}^H}\mbf{H}_{2}\mbf{p}|^2 }
{1+  \rho  \alpha_{1}^2 |{\mbf{v}_{2}^H}\mbf{H}_{2}\mbf{p}|^2} \bigg).
\end{equation}

As for OMA, under any given power coefficients $\alpha_{1'}$ and $\alpha_{2'}$, satisfying $\alpha_{1'}^2+\alpha_{2'}^2=1$, the split of the DoF between the two users is optimized to achieve the maximum sum rate for fair comparison. We use $\lambda_{1}$ and $\lambda_{2}$ to denote the fractions of the DoF for users $1$ and $2$, respectively, which should satisfy $\lambda_{1}+\lambda_{2}=1$. As such, the achievable rate at user $k$ can be expressed as \cite{18}
%$R_{k}^{\text{OMA}}=\lambda_{k} \log_2
%\begin{pmatrix}
%1+ \frac{\rho \alpha_{k}^2 |{\mbf{v}_{k}^H}\mbf{H}_{k}\mbf{p}_k|^2 }
%{\lambda_{k}}
%\end{pmatrix}$. 

\begin{equation}
R_{k}^{\text{OMA}}=\lambda_{k} \log_2
\begin{pmatrix}
1+ \frac{\rho \alpha_{k'}^2 |{\mbf{v}_{k}^H}\mbf{H}_{k}\mbf{p}_k|^2 }
{\lambda_{k}}
\end{pmatrix}.
\end{equation}

Now, the sum rate for the two users in the same cluster in MIMO-OMA is given by \cite[Lemma 1]{18}
\begin{equation} \label{eq:OMA}
{\color{black} R^{\text{OMA}}_{\text{sum}} } \leq \log_2 \bigg(1+  \sum_{k=1}^{2} \rho \alpha_{k'}^2 |{\mbf{v}_{k}^H}\mbf{H}_{k}\mbf{p}_k|^2  \bigg),
\end{equation}
where the equality holds for
\begin{equation} \label{eq:COMA}
{\lambda_{k}=\frac{ \alpha_{k'}^2 |{\mbf{v}_{k}^H}\mbf{H}_{k}\mbf{p}_k|^2 } { \sum_{l=1}^{2}  \alpha_{l'}^2 |{\mbf{v}_{l}^H}\mbf{H}_{l}\mbf{p}_l|^2} }.
\end{equation}

%The following lemma provides the condition for the splits of DoF in order to achieve the capacity for OMA.
%\begin{lemma}
%The sum rate for the two users in the same cluster in MIMO-OMA is given by \cite{18}
%\begin{equation} \label{eq:OMA}
%{\color{black} R^{\text{OMA}}_{\text{sum}} } \leq \log_2 \bigg(1+  \sum_{k=1}^{2} \rho \alpha_{k'}^2 |{\mbf{v}_{k}^H}\mbf{H}_{k}\mbf{p}_k|^2  \bigg),
%\end{equation}
%where the equality holds when the following condition is met:
%\begin{equation} \label{eq:COMA}
%{\lambda_{k}=\frac{ \alpha_{k'}^2 |{\mbf{v}_{k}^H}\mbf{H}_{k}\mbf{p}_k|^2 } { \sum_{l=1}^{2}  \alpha_{l'}^2 |{\mbf{v}_{l}^H}\mbf{H}_{l}\mbf{p}_l|^2} }.
%\end{equation}
%\end{lemma}

%\begin{remark}
Note that when \eqref{eq:COMA} is satisfied, the maximum sum rate for OMA is achieved, and the corresponding individual rates for users 1 and 2 are used for OMA to ensure a fair comparison.
%\end{remark}
%\begin{IEEEproof}
%Refer to \cite{18}.
%\end{IEEEproof}

\subsection{Problem Formulation}
In \cite{18}, the authors prove that NOMA can achieve a larger sum rate than OMA by simply assigning the same power coefficients to both schemes. However, having a higher sum rate does not guarantee that each user in NOMA has a higher data rate than its counterpart in OMA. Indeed, it is easy to come up with an instance in which the data rate of the weak user (user 2) in NOMA is below its counterpart in OMA if simply assigning the same power coefficients. For example, if \text{$\rho \alpha_{1}^2 |{\mbf{v}_{1}^H}\mbf{H}_{1}\mbf{p}|^2 =\rho \alpha_{2}^2 |{\mbf{v}_{2}^H}\mbf{H}_{2}\mbf{p}|^2 =0.25$} and $\alpha_{1}^2=0.5\alpha_{2}^2$, then $\log_2 
\bigg(1+\frac{\rho \alpha_{2}^2 |{\mbf{v}_{2}^H}\mbf{H}_{2}\mbf{p}|^2 }
{1+  \rho  \alpha_{1}^2 |{\mbf{v}_{2}^H}\mbf{H}_{2}\mbf{p}|^2} \bigg)=\log_2 (1.22) 
 < \log_2(1.23) \allowbreak = \frac{\alpha_{2}^2 |{\mbf{v}_{2}^H}\mbf{H}_{2}\mbf{p}|^2}{\alpha_{1}^2 |{\mbf{v}_{1}^H}\mbf{H}_{1}\mbf{p}|^2+\alpha_{2}^2 |{\mbf{v}_{2}^H}\mbf{H}_{2}\mbf{p}|^2}  \log_2 \bigg(1+\rho \alpha_{1}^2 |{\mbf{v}_{1}^H}\mbf{H}_{1}\mbf{p}|^2+   \allowbreak \rho \alpha_{2}^2 |{\mbf{v}_{2}^H}\mbf{H}_{2}\mbf{p}|^2 \bigg) $. This means that NOMA may lead to unfair data rate between its two users when compared with OMA. Consequently, to further verify the superiority of NOMA over OMA, PA should be conducted such that the data rate of each user in NOMA exceeds its counterpart in OMA. A PA scheme satisfying this requirement is proposed in \cite{17}. However, \cite{17} adopts time-division multiple access with equal power and DoF for its users as the representative of OMA, which does not achieve maximum sum rate for OMA. On the other hand, for a general case, like any PA for OMA, does this conclusion still hold? To the best of our knowledge, this problem has never been considered in the literature.

To validate that NOMA achieves a higher individual rate than OMA for an arbitrary PA in OMA, we need to find the feasible power coefficients for NOMA, which achieve this goal under any given power coefficients and optimal DoF for OMA. The considered problem can be formulated as follows:
%\begin{subequations}\label{eq:PA}
%\begin{align}
%    \text{find} ~ & \alpha_{1}, \alpha_{2}\\ 
%\text{subject to }  &(10)\\
%     \log_2(1+ &\rho \alpha_{1}^2 |{\mbf{v}_{1}^H}\mbf{H}_{1}\mbf{p}|^2) 
%     \geq \lambda_1 \log_2 \bigg( 1+\frac{\rho \alpha_{1'}^2 |{\mbf{v}_{1}^H}\mbf{H}_{1}\mbf{p}|^2}{\lambda_1} \bigg), \\
%    \log_2 \bigg(1+ &\frac{\rho \alpha_{2}^2 |{\mbf{v}_{2}^H}\mbf{H}_{2}\mbf{p}|^2 }
%{1+  \rho  \alpha_{1}^2 |{\mbf{v}_{2}^H}\mbf{H}_{2}\mbf{p}|^2} \bigg) \\ \nonumber
%&\geq \lambda_2 \log_2 \bigg( 1+\frac{\rho \alpha_{2'}^2 |{\mbf{v}_{2}^H}\mbf{H}_{2}\mbf{p}|^2}{\lambda_2} \bigg) ,
%\end{align}
%\end{subequations} 
\begin{IEEEeqnarray*}{clr}\label{eq:PA}
\displaystyle {\text{find}}  & \quad \alpha_{1}, \alpha_{2} \IEEEyesnumber \IEEEyessubnumber* \\  
\text{s.t.} &\quad (10), \\ \nonumber
&\quad \log_2(1+ \rho \alpha_{1}^2 |{\mbf{v}_{1}^H}\mbf{H}_{1}\mbf{p}|^2) \\
  &\quad \geq \lambda_1 \log_2 \bigg( 1+\frac{\rho \alpha_{1'}^2 |{\mbf{v}_{1}^H}\mbf{H}_{1}\mbf{p}|^2}{\lambda_1} \bigg),  \\  \nonumber
   &\quad \log_2 \bigg(1+ \frac{\rho \alpha_{2}^2 |{\mbf{v}_{2}^H}\mbf{H}_{2}\mbf{p}|^2 }
{1+  \rho  \alpha_{1}^2 |{\mbf{v}_{2}^H}\mbf{H}_{2}\mbf{p}|^2} \bigg) \\   
&\quad \geq \lambda_2 \log_2 \bigg( 1+\frac{\rho \alpha_{2'}^2 |{\mbf{v}_{2}^H}\mbf{H}_{2}\mbf{p}|^2}{\lambda_2} \bigg), \\  
&\quad \alpha_{1}^2+\alpha_{2}^2=1, \alpha_{1}^2, \alpha_{2}^2 \in [0,1],
\end{IEEEeqnarray*}
where (\ref{eq:PA}b) ensures that OMA achieves the maximum sum rate, while (\ref{eq:PA}c) and (\ref{eq:PA}d) guarantee that NOMA outperforms OMA for both users. 

\section{Proposed PA scheme}
\subsection{Optimal DoF and Varying Power}
In this section, we propose a PA strategy, which satisfies the constraints (\ref{eq:PA}b)-(\ref{eq:PA}e). First, with some algebraic manipulations on (\ref{eq:PA}c) and (\ref{eq:PA}d), the PA strategy for NOMA is given by 
\begin{subequations}\label{eq:PA_con}
\begin{align}
  &\alpha_{1}^2 \geq \frac{(1+\frac{\rho \alpha_{1'}^2 |{\mbf{v}_{1}^H}\mbf{H}_{1}\mbf{p}|^2}{\lambda_1})^ {\lambda_1} -1}{\rho |{\mbf{v}_{1}^H}\mbf{H}_{1}\mbf{p}|^2}, \\
  & \alpha_{1}^2 \leq \frac{1+ \rho |{\mbf{v}_{2}^H}\mbf{H}_{2}\mbf{p}|^2 - (1+\frac{\rho \alpha_{2'}^2 |{\mbf{v}_{2}^H}\mbf{H}_{2}\mbf{p}|^2}{\lambda_2})^ {\lambda_2}}{\rho |{\mbf{v}_{2}^H}\mbf{H}_{2}\mbf{p}|^2 (1+\frac{\rho \alpha_{2'}^2 |{\mbf{v}_{2}^H}\mbf{H}_{2}\mbf{p}|^2}{\lambda_2})^ {\lambda_2}}.
\end{align}
\end{subequations}
%Obviously, $P_1'\geq 0$. 
 
%Note that when $\lambda_1=\lambda_2=1/2$ and $P_1=P_2=1/2$, the above constraint is the same as [(4-5), 6].

Now, to ensure a feasible solution for $\alpha_{1}^2$, the following condition must be satisfied: 
\begin{equation} \label{eq:p_1'}
\begin{split}
&\frac{(1+\frac{\rho \alpha_{1'}^2 |{\mbf{v}_{1}^H}\mbf{H}_{1}\mbf{p}|^2}{\lambda_1})^ {\lambda_1} -1}{\rho |{\mbf{v}_{1}^H}\mbf{H}_{1}\mbf{p}|^2} 
 \\
 &\leq 
 \frac{1+ \rho |{\mbf{v}_{2}^H}\mbf{H}_{2}\mbf{p}|^2 - (1+\frac{\rho \alpha_{2'}^2 |{\mbf{v}_{2}^H}\mbf{H}_{2}\mbf{p}|^2}{\lambda_2})^ {\lambda_2}}{\rho |{\mbf{v}_{2}^H}\mbf{H}_{2}\mbf{p}|^2 (1+\frac{\rho \alpha_{2'}^2 |{\mbf{v}_{2}^H}\mbf{H}_{2}\mbf{p}|^2}{\lambda_2})^ {\lambda_2}}.
 \end{split}
\end{equation}

With the help of (\ref{eq:PA}b), and after some algebraic manipulations, (\ref{eq:p_1'}) can be further expressed as

\begin{IEEEeqnarray*}{llc} \label{eq:p_11'}
&(|{\mbf{v}_{1}^H}\mbf{H}_{1}\mbf{p}|^2-|{\mbf{v}_{2}^H}\mbf{H}_{2}\mbf{p}|^2) 
\times 
\bigg( 
\ub{ 1+\rho \alpha_{2'}^2 |{\mbf{v}_{2}^H}\mbf{H}_{2}\mbf{p}|^2 }_{\text{the first part}}- \notag \\
&\ub{ (1+\rho \alpha_{1'}^2 |{\mbf{v}_{1}^H}\mbf{H}_{1}\mbf{p}|^2+\rho \alpha_{2'}^2 |{\mbf{v}_{2}^H}\mbf{H}_{2}\mbf{p}|^2)^{\frac{\alpha_{2'}^2 |{\mbf{v}_{2}^H}\mbf{H}_{2}\mbf{p}|^2}{\alpha_{1'}^2 |{\mbf{v}_{1}^H}\mbf{H}_{1}\mbf{p}|^2+\alpha_{2'}^2 |{\mbf{v}_{2}^H}\mbf{H}_{2}\mbf{p}|^2}} }_{\text{the second part}} \bigg) \notag \\
&\geq 0.\IEEEyesnumber
\end{IEEEeqnarray*}

In the following lemma, we ensure that \eqref{eq:p_11'} always holds for any PA and optimal DoF for OMA. 

\begin{lemma}
Equation \eqref{eq:p_11'} always holds.
\end{lemma}

\begin{IEEEproof}
Since $|{\mbf{v}_{1}^H}\mbf{H}_{1}\mbf{p}|^2 \geq |{\mbf{v}_{2}^H}\mbf{H}_{2}\mbf{p}|^2$, we only need to show that the second term of \eqref{eq:p_11'} is non-negative. We observe the following:

\begin{itemize}
\item the first part is a linear function over $\alpha_{2'}^2$;

\item the second part is a convex function over $\alpha_{2'}^2$ when $\alpha_{2'}^2 \in [0,1]$; 

%\footnote{ We cannot use derivative to prove the convexity of the second part directly, as $\alpha_{2'}^2$ is in both the bottom and the power. Instead, we use $\log(\cdot)$ to make it possible to calculate the derivative. According to the composite rule of convex functions, the original function is a convex one, if its $\log(\cdot)$ function is a convex one. The $\log(\cdot)$ function has the form of $\frac{x}{1+x} \log(1+x)$, which is convex when $x \in [0,1]$ \cite{4}. As such, the second part is convex. }

\item the first and second parts intersect when $\alpha_{2'}^2=0$ or \text{$\alpha_{2'}^2=1$}.
\end{itemize}

According to the characteristics of convex function, the line segment between any two points on the graph lies above the graph. Thus, the second term of \eqref{eq:p_11'} is always non-negative when $\alpha_{2'}^2 \in [0,1]$. 
\end{IEEEproof}

As a result, we can claim that for any value of $\alpha_{1}^2$ satisfying \eqref{eq:PA_con}, MIMO-NOMA provides higher individual rates when compared with MIMO-OMA.

%\section{Proposed PA strategy}
%In the formulated problem, since \eqref{eq:COMA} is required to maximize the sum rate for MIMO-OMA,

\subsection{Equal DoF and Varying Power}
In \eqref{eq:PA}, the DoF are split according to \eqref{eq:COMA}. As the PA is arbitrary, the resulting fractions of DoF can also take any value, which may be infeasible to realize in practice \cite{23}. Motivated by this observation, in this section, we consider a simple and practical case when the DoF for two users in the same cluster in MIMO-OMA are equal, while the PA is still arbitrary. Compared with \cite{17}, the considered case is more general as the PA can be arbitrary. In contrast to \cite{6}, which only ensures the QoS of the weak user, the considered case takes into account both strong and weak users. 

%Since NOMA will assign all the power to the user with better channel gains if the goal is to maximize sum rate, we assume that the user with worse channel gains has a QoS requirement. On this basis, we maximize the sum rate. For NOMA, the PA should follow that the data rate of the weak user is just satisfied, and the rest is allocated to the strong user. Accordingly, we have the following equations:

The corresponding problem can be formulated as: 
\begin{IEEEeqnarray*}{clr}\label{eq:equal_degree}
\displaystyle {\text{find}}  & \quad \alpha_{1}, \alpha_{2} \IEEEyesnumber \IEEEyessubnumber* \\  \nonumber
\text{s.t.} 
&\quad \log_2(1+ \rho \alpha_{1}^2 |{\mbf{v}_{1}^H}\mbf{H}_{1}\mbf{p}|^2) \\
&\quad \geq \frac{1}{2} \log_2 ( 1+2 \rho \alpha_{1'}^2 |{\mbf{v}_{1}^H}\mbf{H}_{1}\mbf{p}|^2 ) ,  \\  \nonumber
   &\quad \log_2 \bigg(1+ \frac{\rho \alpha_{2}^2 |{\mbf{v}_{2}^H}\mbf{H}_{2}\mbf{p}|^2 }
{1+  \rho  \alpha_{1}^2 |{\mbf{v}_{2}^H}\mbf{H}_{2}\mbf{p}|^2} \bigg) \\
&\quad \geq  \frac{1}{2} \log_2 ( 1+2 \rho \alpha_{2'}^2 |{\mbf{v}_{2}^H}\mbf{H}_{2}\mbf{p}|^2 ), \\  
&\quad \alpha_{1}^2+\alpha_{2}^2=1, \alpha_{1}^2, \alpha_{2}^2 \in [0,1].
\end{IEEEeqnarray*}
%\begin{figure}
%\centering
%\includegraphics[width=0.5\textwidth]{eps/fig1-2.eps}
%\caption{Numerical study to show that NOMA achieves higher individual data rate than OMA as the power coefficient varies.}
%\end{figure}

Note that the main difference between \eqref{eq:equal_degree} and \eqref{eq:PA} lies in the fact that (\ref{eq:PA}b) is no longer a constraint in the former. Instead, both $\lambda_1$ and $\lambda_2$ take a fixed value of $\frac{1}{2}$. 

To find the solution of \eqref{eq:equal_degree}, we start with the case when equality is attained in (\ref{eq:equal_degree}c). Accordingly, we have
\begin{subequations}\label{eq:weak_user_equal}
\begin{align} 
\frac{(1+\rho  |{\mbf{v}_{2}^H}\mbf{H}_{2}\mbf{p}|^2)^2}{(1+\rho \alpha_{1}^2 |{\mbf{v}_{2}^H}\mbf{H}_{2}\mbf{p}|^2)^2}=1+2\rho (1-\alpha_{1'}^2) |{\mbf{v}_{2}^H}\mbf{H}_{2}\mbf{p}|^2 \\
\Longleftrightarrow \alpha_{1}^2= 
\frac{1+\rho |{\mbf{v}_{2}^H}\mbf{H}_{2}\mbf{p}|^2 - \sqrt{1+ 2\rho \alpha_{2'}^2 |{\mbf{v}_{2}^H}\mbf{H}_{2}\mbf{p}|^2}}{\rho |{\mbf{v}_{2}^H}\mbf{H}_{2}\mbf{p}|^2 \sqrt{1+ 2\rho \alpha_{2'}^2 |{\mbf{v}_{2}^H}\mbf{H}_{2}\mbf{p}|^2}}.
\end{align}
\end{subequations}

On this basis, we ensure that (\ref{eq:equal_degree}b) always holds. To achieve that, we rewrite (\ref{eq:equal_degree}b) as
\begin{equation} \label{eq:weak_equal1}
(1+\rho \alpha_{1}^2 |{\mbf{v}_{1}^H}\mbf{H}_{1}\mbf{p}|^2)^2 \geq  1+2 \rho \alpha_{1'}^2 |{\mbf{v}_{1}^H}\mbf{H}_{1}\mbf{p}|^2,
\end{equation}
and (\ref{eq:weak_user_equal}a) as
%\begin{equation} \label{eq:weak_proof}
%\begin{split}
\begin{IEEEeqnarray*}{lll} \label{eq:weak_proof}
&1+2 \rho \alpha_{1'}^2 |{\mbf{v}_{2}^H}\mbf{H}_{2}\mbf{p}|^2  \\ 
&=
2(1+ \rho  |{\mbf{v}_{2}^H}\mbf{H}_{2}\mbf{p}|^2 )  
- \frac{(1+\rho  |{\mbf{v}_{2}^H}\mbf{H}_{2}\mbf{p}|^2)^2}{(1+\rho \alpha_{1}^2 |{\mbf{v}_{2}^H}\mbf{H}_{2}\mbf{p}|^2)^2} \\  
& =
2(1+ \rho  |{\mbf{v}_{2}^H}\mbf{H}_{2}\mbf{p}|^2 ) + 
(1+\rho \alpha_{1}^2 |{\mbf{v}_{2}^H}\mbf{H}_{2}\mbf{p}|^2)^2  \\ 
&\quad - \bigg [ (1+\rho \alpha_{1}^2 |{\mbf{v}_{2}^H}\mbf{H}_{2}\mbf{p}|^2)^2
+ \frac{(1+\rho  |{\mbf{v}_{2}^H}\mbf{H}_{2}\mbf{p}|^2)^2}{(1+\rho \alpha_{1}^2 |{\mbf{v}_{2}^H}\mbf{H}_{2}\mbf{p}|^2)^2} \bigg] \\ 
& \leq 
2(1+ \rho  |{\mbf{v}_{2}^H}\mbf{H}_{2}\mbf{p}|^2 ) + 
(1+\rho \alpha_{1}^2 |{\mbf{v}_{2}^H}\mbf{H}_{2}\mbf{p}|^2)^2  \\ 
&\quad - 2 (1+\rho  |{\mbf{v}_{2}^H}\mbf{H}_{2}\mbf{p}|^2) \\ 
&= (1+\rho \alpha_{1}^2 |{\mbf{v}_{2}^H}\mbf{H}_{2}\mbf{p}|^2)^2,  \IEEEyesnumber
\end{IEEEeqnarray*}
%\end{split}
%\end{equation}
where the inequality comes from the Jensen's inequality.

%\begin{align} \nonumber
%&1+2 \rho \alpha_{1'}^2 |{\mbf{v}_{2}^H}\mbf{H}_{2}\mbf{p}|^2  \\ 
%&=
%2(1+ \rho  |{\mbf{v}_{2}^H}\mbf{H}_{2}\mbf{p}|^2 )  
%- \frac{(1+\rho  |{\mbf{v}_{2}^H}\mbf{H}_{2}\mbf{p}|^2)^2}{(1+\rho \alpha_{1}^2 |{\mbf{v}_{2}^H}\mbf{H}_{2}\mbf{p}|^2)^2} \\ \nonumber 
%& =
%2(1+ \rho  |{\mbf{v}_{2}^H}\mbf{H}_{2}\mbf{p}|^2 ) + 
%(1+\rho \alpha_{1}^2 |{\mbf{v}_{2}^H}\mbf{H}_{2}\mbf{p}|^2)^2  \\ \nonumber
%&\quad - \bigg [ (1+\rho \alpha_{1}^2 |{\mbf{v}_{2}^H}\mbf{H}_{2}\mbf{p}|^2)^2
%+ \frac{(1+\rho  |{\mbf{v}_{2}^H}\mbf{H}_{2}\mbf{p}|^2)^2}{(1+\rho \alpha_{1}^2 |{\mbf{v}_{2}^H}\mbf{H}_{2}\mbf{p}|^2)^2} \bigg] \\ \nonumber
%& \leq 
%2(1+ \rho  |{\mbf{v}_{2}^H}\mbf{H}_{2}\mbf{p}|^2 ) + 
%(1+\rho \alpha_{1}^2 |{\mbf{v}_{2}^H}\mbf{H}_{2}\mbf{p}|^2)^2  \\ \nonumber
%&\quad - 2 (1+\rho  |{\mbf{v}_{2}^H}\mbf{H}_{2}\mbf{p}|^2)^2 \\ \nonumber
%&= (1+\rho \alpha_{1}^2 |{\mbf{v}_{2}^H}\mbf{H}_{2}\mbf{p}|^2)^2,
%\end{align}
%\end{subequations}

Now, with the help of \eqref{eq:order} and \eqref{eq:weak_proof}, we obtain
\begin{equation}
\begin{split}
&(1+\rho \alpha_{1}^2 |{\mbf{v}_{1}^H}\mbf{H}_{1}\mbf{p}|^2)^2 - 1-2 \rho \alpha_{1'}^2 |{\mbf{v}_{1}^H}\mbf{H}_{1}\mbf{p}|^2 \\
&\geq (1+\rho \alpha_{1}^2 |{\mbf{v}_{2}^H}\mbf{H}_{2}\mbf{p}|^2)^2 - 1-2 \rho \alpha_{1'}^2 |{\mbf{v}_{2}^H}\mbf{H}_{2}\mbf{p}|^2  \\
&\geq 0,
\end{split}
\end{equation}
which is exactly (\ref{eq:weak_equal1}). Hence, (\ref{eq:equal_degree}b) always holds.
%On the other hand, we have
%\begin{subequations}\label{eq:weak_proof1}
%\begin{align}
%&\frac{(1+\rho  |{\mbf{v}_{2}^H}\mbf{H}_{2}\mbf{p}|^2)^2}{(1+\rho \alpha_{1}^2 |{\mbf{v}_{2}^H}\mbf{H}_{2}\mbf{p}|^2)^2} 
%+
%(1+\rho \alpha_{1}^2 |{\mbf{v}_{1}^H}\mbf{H}_{1}\mbf{p}|^2)^2 \\ \nonumber
%&\geq
%2 (1+\rho  |{\mbf{v}_{2}^H}\mbf{H}_{2}\mbf{p}|^2)^2 \\
%\Rightarrow  &
%2 (1+\rho  |{\mbf{v}_{2}^H}\mbf{H}_{2}\mbf{p}|^2)^2
%-
%\frac{(1+\rho  |{\mbf{v}_{2}^H}\mbf{H}_{2}\mbf{p}|^2)^2}{(1+\rho \alpha_{1}^2 |{\mbf{v}_{2}^H}\mbf{H}_{2}\mbf{p}|^2)^2} \\ \nonumber
%&\leq 
%(1+\rho \alpha_{1}^2 |{\mbf{v}_{1}^H}\mbf{H}_{1}\mbf{p}|^2)^2 \\
%\Rightarrow &
%1+2 \rho \alpha_{1'}^2 |{\mbf{v}_{1}^H}\mbf{H}_{1}\mbf{p}|^2
%\leq 
%(1+\rho \alpha_{1}^2 |{\mbf{v}_{1}^H}\mbf{H}_{1}\mbf{p}|^2)^2,
%\end{align}
%\end{subequations} 
%Thus, we have proved (\ref{eq:weak_equal1}), which means that (\ref{eq:equal_degree}b) holds. 

Similarly, we can prove that when equality is achieved for (\ref{eq:equal_degree}b), (\ref{eq:equal_degree}c) holds.  In this case, we have the PA strategy for NOMA as
\begin{equation} \label{eq:value_1}
\alpha_{1}^2=\frac{\sqrt{1+2 \rho \alpha_{1'}^2 |{\mbf{v}_{1}^H}\mbf{H}_{1}\mbf{p}|^2}-1}{\rho  |{\mbf{v}_{1}^H}\mbf{H}_{1}\mbf{p}|^2}.
\end{equation} 

Clearly, when $\alpha_{1}^2$ lies in the boundary between the values in \eqref{eq:weak_user_equal} and \eqref{eq:value_1}, MIMO-NOMA always achieves higher individual rates than MIMO-OMA.

\begin{figure}
\centering
\includegraphics[width=0.5\textwidth]{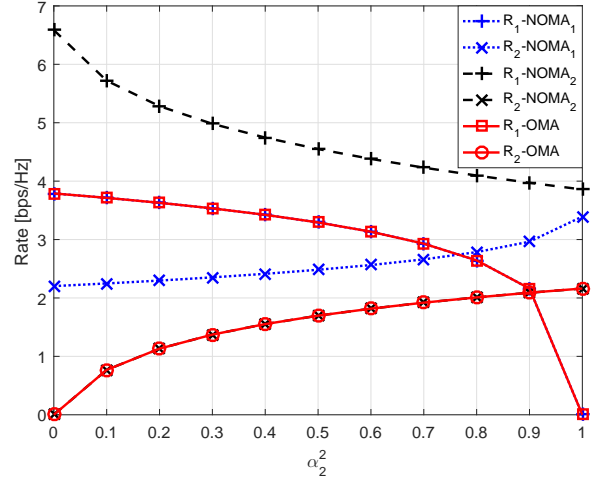}
\caption{Individual rate comparison between NOMA and OMA with equal DoF, as the power coefficient of user 2 for OMA varies.}
\end{figure}

\section{Simulation Results}
In this section, simulations are conducted to compare the individual rates of MIMO-NOMA and MIMO-OMA, and hence, verify the accuracy of the developed analytical results. In simulations, $M=4$ and the path-loss exponent is 3.8.

Fig. 1 compares the individual rate between MIMO-NOMA and MIMO-OMA with equal DoF, when the power coefficient for the weak user varies. In simulations, \text{$\rho=30$ dB}, $|{\mbf{v}_{1}^H}\mbf{H}_{1}\mbf{p}|^2=0.052$ and $|{\mbf{v}_{2}^H}\mbf{H}_{2}\mbf{p}|^2=0.0052$. Note that $\text{NOMA}_1$ and $\text{NOMA}_2$ denote the cases when the power coefficient of the strong user in MIMO-NOMA satisfies \eqref{eq:value_1} and (\ref{eq:weak_user_equal}b), respectively. As expected, $R_1$ in $\text{NOMA}_1$ equals  that in OMA, while $R_2$ in $\text{NOMA}_2$ is the same as that in OMA. Moreover, both $R_2$ in $\text{NOMA}_1$ and $R_1$ in $\text{NOMA}_2$ exceed  their counterparts in OMA, which verifies the superiority of NOMA over OMA in terms of individual rate. Particularly, when $\alpha_{2'}^2 \in [0,0.8]$, it can be seen that $R_1^{\text{NOMA}_1}=R_1^{\text{OMA}} > R_2^{\text{NOMA}_1} > R_2^{\text{OMA}} $. Therefore, $\text{NOMA}_1$ also provides better fairness than OMA. 

\begin{figure}
\centering
\includegraphics[width=0.5\textwidth]{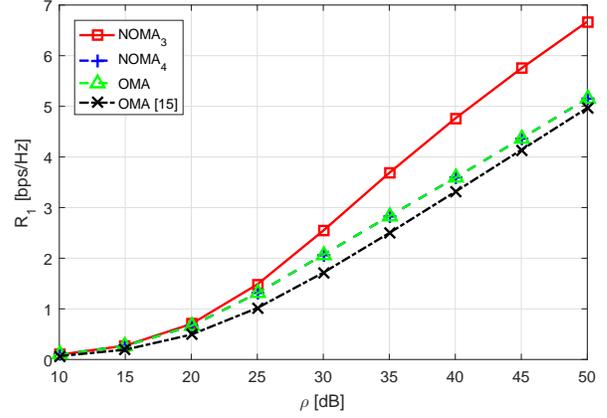}
\caption{The rate of user 1, i.e., $R_1$ versus $\rho$ for both NOMA and OMA.}
\end{figure}

%Figures 2-4 present results obtained when the optimal DoF is used for OMA. Note that the legends U-NOMA and \text{L-NOMA} in Figs. 2 and 4 denote the scenarios when $\alpha_{1}^2$ satisfies (\ref{eq:PA_con}b) and (\ref{eq:PA_con}a), respectively. In contrast, in \text{Fig. 3}, the legends U-NOMA and L-NOMA means the opposite order, i.e., when $\alpha_{1}^2$ satiesfies (\ref{eq:PA_con}a) and (\ref{eq:PA_con}b), respectively. In addition, the legends OMA \cite{17} and OMA denote the OMA scheme in \cite{17} (with equal power and DoF) and the one considered in this paper (with arbitrary power and optimal DoF), respectively. 

Figs. 2-4 present results obtained when the optimal DoF is used for OMA. The legends $\text{NOMA}_3$ and $\text{NOMA}_4$ denote the scenarios when $\alpha_{1}^2$ follows (\ref{eq:PA_con}b) and (\ref{eq:PA_con}a), respectively. In addition, the legends OMA \cite{17} and OMA denote the OMA scheme in \cite{17} (with equal power and DoF) and the one considered in this paper (with arbitrary power and optimal DoF), respectively. 

\begin{figure}
\centering
\includegraphics[width=0.5\textwidth]{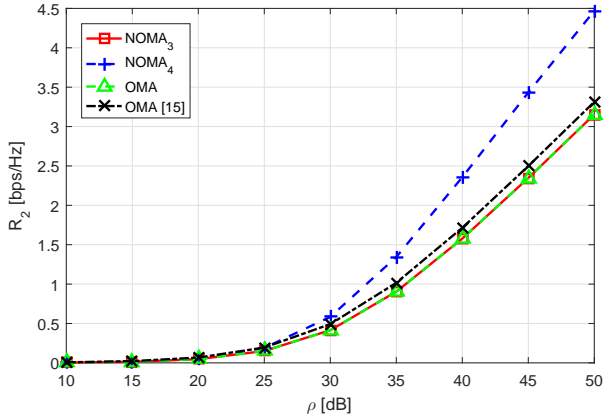}
\caption{The rate of user 2, i.e., $R_2$ versus $\rho$ for both NOMA and OMA.}
\end{figure}

%In Fig. 2, we show how $R_1$ for the above four schemes varies with $\rho$. It can be seen that U-NOMA achieves the highest rate for $R_1$, while OMA in \cite{17} obtains the lowest rate. In addition, L-NOMA attains the same rate as OMA considered in this paper. Likewise, in Fig. 3, we illustrate how $R_2$ for the above four schemes varies with $\rho$. Clearly, \text{U-NOMA} achieves the highest rate for $R_2$, being followed by OMA \cite{17}. L-NOMA obtains the same rate as OMA. Combining these two figures, we can easily conclude that NOMA can always achieve higher individual rates than OMA considered in this paper, once \eqref{eq:PA_con} is satisfied. Furthermore, when L-NOMA in Fig. 2 is used, better fairness is achieved for NOMA than OMA. On the other hand, NOMA also outperforms OMA \cite{17} for both $R_1$ and $R_2$, when equality is obtained in (\ref{eq:PA_con}a) for $\alpha_{1}^2$. In this case, $R_2$ for U-NOMA is higher than that for OMA \cite{17}. Also, $R_1$ in NOMA is better than that in OMA in \cite{17}. 
%
%
%
%Lastly, from Fig. 4, we can observe that OMA in this paper has a larger sum rate than OMA \cite{17} owing to use of optimal DoF. This justifies the necessity of optimizing the DoF for the comparison between NOMA and OMA. The order of the sum rate is $\text{U-NOMA}>\text{L-NOMA}>\text{OMA}>\text{OMA \cite{17}}$. 

In Fig. 2, we show how $R_1$ varies with $\rho$ for the previously mentioned four schemes. It can be seen that $\text{NOMA}_3$ achieves the highest rate for $R_1$, while OMA in \cite{17} obtains the lowest rate. In addition, $\text{NOMA}_4$ attains the same rate as OMA considered in this paper. Likewise, in Fig. 3, we illustrate how $R_2$ varies with $\rho$ for the above four schemes. Clearly, $\text{NOMA}_4$ achieves the highest rate for $R_2$, being followed by OMA \cite{17}. $\text{NOMA}_3$ obtains the same rate as OMA. Combining these two figures, we can easily conclude that NOMA can always achieve higher individual rates than OMA considered in this paper, once \eqref{eq:PA_con} is satisfied. Particularly, under $\text{NOMA}_4$, better fairness is achieved by NOMA when compared with OMA. Morover, NOMA also outperforms OMA \cite{17}, as both $R_1$ and  $R_2$ for $\text{NOMA}_4$ are higher than that for OMA \cite{17}.

Lastly, from Fig. 4, we can observe that OMA considered in this paper has a larger sum rate than OMA \cite{17} owing to the use of optimal DoF. This justifies the necessity of optimizing the DoF for the comparison between NOMA and OMA. The order of the sum rate is $\text{NOMA}_3>\text{NOMA}_4>\text{OMA}>\text{OMA \cite{17}}$. $\text{NOMA}_3>\text{NOMA}_4$ can be explained by the fact that allocating more power to the stronger user results in a higher sum rate.

\section{Conclusion}
A fair individual rate comparison between MIMO-NOMA and MIMO-OMA has been investigated. We have proposed a PA strategy, which guarantees that MIMO-NOMA achieves a higher individual rate than MIMO-OMA with arbitrary power coefficients and optimized DoF. Additionally, we have shown that this also holds for the case of equal DoF and arbitrary power coefficients. Numerical results verify the accuracy of the developed analytical results.

\begin{figure}
\centering
\includegraphics[width=0.5\textwidth]{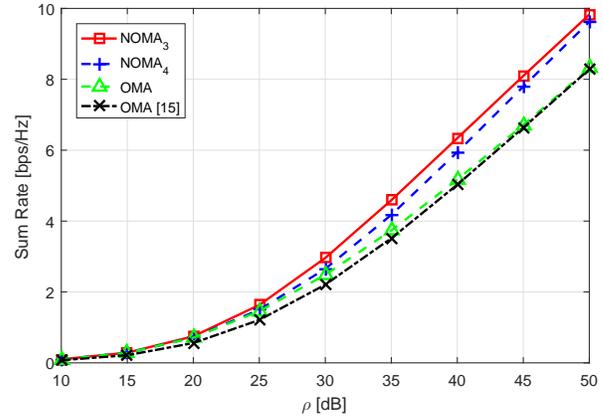}
\caption{The sum rate versus $\rho$ for both NOMA and OMA.}
\end{figure}
%In this paper, we have proposed to conduct capacity comparison in a fair manner. Due to most of the existing literature does not maximize the sum rate of OMA, we first give the condition on how to satisfy this requirement. On this basis, we have prove that NOMA still outperforms OMA for both SISO and MIMO case. Moreover, we have validated that each user in NOMA can obtain a larger data rate than its counterpart through appropriate PA. 

%\bibliographystyle{IEEEtran}
%%\bibliography{IEEEabrv,NOMA-OMA Comparison}
%
%\bibliography{mybibfile}
\bibliographystyle{IEEEtran}
\bibliography{IEEEabrv,conf_short,jour_short,mybibfile}

\end{document}